\documentstyle[aps,pre,psfig,epsfig,12pt]{revtex}
\begin{document}
\bibliographystyle{unsrt}

\draft
\title{
        Rectification of current in ac-driven nonlinear systems and \\
        symmetry properties of the Boltzmann equation
      }
\author{
        O. Yevtushenko $^1$, S. Flach $^1$, Y. Zolotaryuk $^1$ and
        A. A. Ovchinnikov $^{1,2}$
       }       
\address{
$^1$ Max--Planck--Institut f\"{u}r Physik komplexer Systeme,
        N\"{o}thnitzer Str. 38, D-01187, \\
	Dresden, Germany \\
$^2$Institute for Chemical Physics of Russian Academy of Sciences, 117977, Moscow, Russia
        }
\date{\today}
%

\maketitle

\begin{abstract}
We study rectification of a current of particles moving in a spatially
periodic potential
under the influence of time-periodic forces with zero mean value.
If certain time-space symmetries are broken a non-zero directed current
of
particles is possible. We investigate this phenomenon in the framework
of the
kinetic Boltzmann equation. We find that the
attractor
of the Boltzmann equation completely reflects the symmetries of the
original one-particle
equation of motion. Especially, we analyse the limits
of weak and strong relaxation. 
The dc current increases by several orders of magnitude with
decreasing dissipation.
\end{abstract}



In recent years, a lot of efforts have been devoted to the study of a directed
current of particles in a spatially periodic potential under the
simultaneous action of
an external time-dependent field with zero mean \cite{hb96-jap97}. Most of the
previously done work  was connected with transport
in the presence of a 
stochastic external field (noise) \cite{b87zp,bhk94,Mark}. Another
set of problems concerns systems driven by a deterministic
periodic force \cite{fyz00prl,m00prl,yfr00pre} with applications to
current generation in semiconductor superlattices \cite{ackkc98},
a free particle
moving in a non-Newtonian liquid \cite{vs85pla}, two-dimensional conducting
electron gas in arrays of triangular shaped quantum dots \cite{lsl99},
to name a few. For more information we refer the interested reader
to \cite{pr00}.

The model Hamiltonian of a particle of unit mass and its
dynamical equation of motion can be written as:
   \begin{equation} \label{Ham}
      H = \frac{p^2}{2} + U(x) - x E(t)\;\;,\;\;
      \ddot{x}=-U'(x) + E(t)\;\;.
   \end{equation}
Here $U$ is a spatially periodic potential $ \, U (x)=U(x+2\pi) \, $,
$E$ is
an  ac field $ \, E(t)=E(t+T) \, $ with zero mean value and frequency
$\omega=2\pi/T$.

A recent approach to the problem of current
rectification \cite{fyz00prl}
was based on the analysis of symmetries of the dynamical equations of
motion
with and without additional dissipation. Consider trajectories
generated from an ensemble of initial conditions in
phase space. A dc
current
can be calculated by averaging over initial conditions 
and time. To generate a non-zero dc current it is necessary
to break all  
relevant symmetries
which are responsible for transforming a trajectory with $\;
x(t;x_0,p_0),~p(t;x_0,p_0) \;$,  
$\; x(t_0;x_0,p_0) = x_0 \;$ and $\; p(t_0;x_0,p_0)=p_0 \; $
into another one with a momentum of the opposite sign. Note, that the
generated trajectory must belong to the same statistical ensemble.
These symmetries can be expressed via combinations of
shifts and reflections in time and space (see \cite{fyz00prl}):
\begin{eqnarray}
&&{\bf {\hat S}_a} \left [ \begin{array}{c} x(t;x_0,p_0)\\ p(t;x_0,p_0)
\end{array} \right]=\left [ \begin{array}{c} -x \left (t+T/2;x_0,p_0
\right )
+2 {\cal {X}} \\ -p\left (t+T/2;x_0,p_0 \right ) \end{array} \right]~,
\label{Sa} \\
&&{\bf {\hat S}_b} \left [ \begin{array}{c} x(t;x_0,p_0)\\ p(t;x_0,p_0)
\end{array} \right]= \left [ \begin{array}{c} x(-t+2\tau;x_0,p_0)\\
-p(-t+2\tau;x_0,p_0) \end{array} \right ]~. \label{Sb}
\end{eqnarray}
Here the constants ${\cal {X}}$ and $\tau$ are defined by the shape of 
functions $U(x)$ and $E(t)$ and are in correspondence with the discrete 
symmetry groups
of the Hamiltonian (\ref{Ham}).
If $U'(x+{\cal {X}})=-U'(-x+{\cal {X}})$ and $E(t)=-E(t+T/2)$ the
dynamical
equations of motion are invariant under the symmetry operation ${\bf
{\hat S}_a}$ (this symmetry has been also observed and discussed
by Ajdari et al \cite{ampp94} for the overdamped case).
On the other hand, if $E(t+\tau)=E(-t+\tau)$, they have ${\bf {\hat
S}_b}$
symmetry (dissipationless case provided). Both symmetries
(\ref{Sa}-\ref{Sb}) may be broken
by choosing appropriate functions $U(x)$ 
and $E(t)$.

In the overdamped case the corresponding dynamical equation
\begin{equation}
\gamma \dot {x}+U'(x)+E(t)=0~,
\end{equation}
may possess an additional symmetry
\begin{eqnarray}
&&{\bf {\hat S}_c} \left [ \begin{array}{c} x(t;x_0,p_0)\\ p(t;x_0,p_0)
\end{array} \right]=\left [ \begin{array}{c} x \left (-t+\tau;x_0,p_0
\right )
+ {\cal {X}} \\ -p\left (-t+\tau;x_0,p_0 \right ) \end{array} \right]~,
\label{Sc} 
\end{eqnarray}
provided that
$U(x)$ and $E(t)$ possess
the following properties: $U'(x)=-U'(x+{\cal {X}})$ and
$E(t)=-E(-t+\tau)$.
Note that the symmetry ${\bf {\hat S}_c}$ if realized is based
on time reversal in the overdamped limit, which is quite unexpected.
An independent discovery of this symmetry has been done 
in \cite{pr00} (coined ``supersymmetry'').

The model analysis in \cite{fyz00prl} 
does not account for more realistic statistical properties of an
ensemble of particles
(like thermalization of the distribution). Still it
provides one
with a
good intuitive understanding of current rectification. On the other
hand, a lot of
solid state applications require a more rigorous statistical description
of transport properties.
The aim of the present paper is to show that the symmetry approach can
be
applied as well to the classical kinetic Boltzmann equation. We will
show in particular
that the attractor of the Boltzmann equation completely reflects the
symmetries of the
original equation of motion for one particle. The kinetic Boltzmann
equation reads (see, for example, \cite{landau})
\begin{eqnarray}
     && \hat{{\cal L}} f \equiv
      \partial_t f + \dot{x} \partial_x f +
 \label{KE-1}                      \dot{p} \partial_p f = {\cal J}(f,F)
~, \\
     && \dot {x}=p~,~~\dot{p}=-U'(x)+E(t) \label{KE-2} ~.
\end{eqnarray}
Here $f \equiv f(t,x,p) $ is an unknown distribution function,  $F
\equiv F(x,p) $
is some equilibrium distribution function chosen as:
\begin{equation}
F(x,p)\equiv F_x(x)F_p(p)=\frac{e^{-p^2/2}}{\sqrt{2\pi}}
\frac{e^{-U(x)}}{L}\;\;,\;\;
L=\int_0^{2\pi}e^{-U(x)} dx\;\;,
\label{EQD}
\end{equation}
and ${\cal J}(f,F)$ is a collision integral. Here we apply
the momentum independent single relaxation time approximation
(also known as ``$\tau$-approximation'') \cite{landau}.
Thus, the collision integral should be written as
   \begin{equation}
      {\cal J}(f,F) = - \nu (f-F) \,
   \end{equation}
where $\nu$ is the dissipation constant or the characteristic relaxation
frequency ($1/\nu$ is the relaxation time
to the equilibrium state).
This is the simplest form
of a collision integral which includes relaxation to the equilibrium
distribution and breaks time-reversal symmetry. We do not intend
to discuss its validity for specific physical realizations.
Rather we consider the resulting kinetic equation as a treatable
case which allows us to demonstrate the qualitative features
of dynamical symmetry breaking. Note that for any finite $\nu$
time reversal symmetry is broken, so that we are concerned with
${\bf \hat{S}_a}$ only. The limit of $\nu \rightarrow 0$ will
lead to a restoring of time reversal symmetry (see below) and 
consequently to a necessary consideration of ${\bf \hat{S}_b}$
as well. Interestingly the opposite limit $\nu \rightarrow \infty$
will also induce a symmetry ${\bf \hat{S}_c}$.   

We start with a perturbative description of the overdamped case 
when $1/\nu$ is the smallest time scale 
in the system (formally, we put $ \, \nu \gg 1 $).
We rewrite Eq. (\ref{KE-1}) in the following form
 \begin{equation} \label{LargeNu}
      f(x,p,t) = F(x,p) - \frac{1}{\nu} \hat{{\cal L}} f(x,p,t) \, .
 \end{equation}
Since $1/\nu$ is a small parameter here, we expand the unknown
function $f$ into a series in terms of $ \, 1/\nu $:
 \begin{equation}
      f(x,p,t) = F(x,p) + \sum_{i=1}^{\infty} (-1)^{i} \frac{ f_i(x,p,t)
}{ \nu^i }~.
 \end{equation}
Collecting terms of the same order, we successively find the
increments $ \, f_i $.
The average value of the dc current is calculated using
\begin{equation}
    j_{dc} = \langle j(t) \rangle_t~,~~j(t)=\langle p f(t,x,p)
\rangle_{x,p} ~,
\label{jdc}
\end{equation}
where $\left <...\right >_{x,p,t}$ stands for averaging over the phase
space coordinates $(x,p)$ and time
correspondingly. We expect the attractor of the Boltzmann equation 
to possess the same
symmetries as the
corresponding dynamical equations of motion, which should be reflected in
the
vanishing or nonvanishing of $j_{dc}$. 
In other words, if the relevant symmetries are broken 
in the equations of motion \cite{fyz00prl}, we will observe this
fact by analysing the increments $f_i$. 
We show this using the following model
\begin{eqnarray}
\label{U} U(x)&=&U_0\left \{ 1-\cos x + \frac{v_2}{2} [1-\cos
(2x+\theta)]
\right \}~, \\
\label{E} E(t)&=&E_1\cos \omega t + E_2 \cos(2\omega t + \alpha) ~.
\end{eqnarray}
As it was already shown in the previous work \cite{fyz00prl}, this
choice of the potential and external force can break both symmetries
(\ref{Sa})-(\ref{Sb}).
We apply the perturbation analysis  for two
specific cases:
(i) $v_2 \neq 0$, $E_2 = 0$ and (ii) $v_2 = 0$, $E_2 \neq 0$.
A related overdamped case was discussed in terms of harmonic
mixing in \cite{sm78ssc-brvw82apb-wb84zpb}.
In both cases the appropriate time-space symmetries are
broken except for specific values of $\theta$ and $\alpha$. 
As already explained, we need 
to break only ${\bf {\hat S}_a}$ since
time reversal and thus ${\bf {\hat S}_b}$ are  
already violated for nonzero $\nu$.
We find (a more detailed description
of these calculations will be reported elsewhere) that the 
lowest order increment
which yields a
non-zero dc current in case (i)
is proportional to $1/\nu^5$ and to the square of the
external field $E(t)$:
\begin{equation}
j_{dc} =-\frac{3}{\nu^5 } \left < E^2(t) \right >_t
\left < U'(x) F_x''(x) \right >_x
 =- \frac{3}{2\nu^5 L}E_1^2 \int_0^{2 \pi}U'''(x) e^{-U(x)}dx\;\;.
\end{equation}
The integral in the above formula cannot be computed analytically.
Yet we find that
it is non-zero only when $v_2 \neq 0$
and $\theta \neq 0,\pi$ because otherwise the symmetry of the potential
function enforces the vanishing of the above expression.
For the case (ii) this term equals zero and the first non-zero contribution
to the current appears in seventh order of the perturbation theory.
As a result we obtain the following expression for the dc current
\begin{equation}
j_{dc} =-\frac{15}{\nu^7 } \left < E^3(t) \right >_t
\left < U'(x) F_x'''(x)\right >_x 
= -\frac{45 }{4\nu^7 } \frac{I_1(U_0)}{I_0(U_0)} E_2E_1^2 \cos
{\alpha}\;\;.
\end{equation}
Here $I_n(z)$ is the modified Bessel function of $n$th order \cite{Abr}.
Note that the averaged current is proportional to $\cos
{\alpha}$ in this case. 
The symmetry which causes the vanishing of the dc current
for $\alpha = \pm \pi/2$ is ${\bf {\hat S}_c}$ [see Eq. (\ref{Sc})].
The above considered case (ii) with $\alpha = \pm \pi /2$
satisfies exactly this symmetry. 
To conclude the consideration of $\nu \gg 1$ we find perfect agreement
between the symmetry properties of the dynamical equations of motion
and the symmetry properties of the used kinetic equation, resulting
in a vanishing or appearance of a dc current.

Let us now investigate 
the case of intermediate $\nu \sim 1$ and weak $\nu \ll 1$ dissipation.
When the characteristic relaxation frequency  $
\, \nu \, $ approaches zero the solution of the kinetic equation
is expected to adequately describe the properties of the 
(microcanonical) ensemble
of trajectories in the phase space of the underlying dynamical system.
Especially we expect that if allowed by the choice of functions
$U(x)$ and $E(t)$, the symmetry ${\bf \hat{S}_b}$ described in detail
in \cite{fyz00prl} should be restored on the
attractor of the kinetic equation
in the limit of $ \, \nu \to 0 $.

The limit of small dissipation appears to be difficult 
to be treated analytically.
We have used numerical
simulations to
calculate the dc current in this case. Two independent numerical
methods have been used.

The first one (method I) is based on the
expansion of the unknown distribution function $f(t,x,p)$ into Hermitian
polynomials
in $ \, p \, $ and Fourier series in $x$-space.
First we expand $f$ into a series w.r.t. $p$:
\begin{equation}
f(t,x,p)=\sum_{n=0}^{\infty} C_n(t,x) |n\rangle
\;\;,\;\;
|n\rangle =\frac{H_n(p)}{\pi ^{1/4} \sqrt{2^n n!}} {\rm e}^{-p^2/2}
\end{equation}
where $H_n(p)$ are Hermite polynomials of $n$-th order.
Since $U(x)$ and $F(x,p)$ are periodic functions in $x$ we may use
\begin{equation}
F(x,p)=\frac{|0\rangle}{2^{1/2}\pi^{1/4}} \sum_{m=-\infty}^{\infty} 
F_m {\rm e}^{imx}
\;\;,\;\;
U(x)=\sum_{m=-\infty}^{\infty} U_m  {\rm e}^{imx}
\;\;,\;\;
C_n(t,x) = \sum_{m=-\infty}^{\infty} A_{n,m}(t) {\rm e}^{imx}\;\;.
\end{equation}
Using $\langle n|n'\rangle = \delta_{n,n'}$ we
arrive at the following infinite set of coupled ordinary differential
equations for $n=0$ 
\begin{equation}
\dot{A}_{0,m} + \frac{1}{\sqrt{2}}E(t)A_{1,m} -\frac{i}{\sqrt{2}}
\sum_{l} l U_{l} A_{n,m-l} + 
\frac{i}{\sqrt{2}}m A_{1,m} =
-\nu (A_{0,m} - F_m)
\end{equation}
and $n > 0$
\begin{eqnarray}
\dot{A}_{n,m} + \frac{1}{\sqrt{2}}E(t) (\sqrt{n+1} A_{n+1,m}
-\sqrt{n} A_{n-1,m})
+ 
\frac{i}{\sqrt{2}} m(\sqrt{n+1} A_{n+1,m} + \sqrt{n}A_{n-1,m})
-
\nonumber \\
\frac{i}{\sqrt{2}}\sqrt{n+1} \sum_l l U_l A_{n+1,m-l}
+ 
\frac{i}{\sqrt{2}}\sqrt{n} \sum_l l U_l A_{n-1,m-l}
= -\nu A_{n,m}\;\;.
\end{eqnarray}
The resulting ODEs for the time-dependent coefficients $A_{n,m}(t)$
are integrated
numerically. After a sufficiently large integration time $\gg 1/\nu$ 
the system
will relax into a time-periodic attractor. On this attractor the
induced current is given by
\begin{equation}
\langle \;p\; \rangle_{p,x} =
3 \sqrt{2} \pi ^{1/4} \sum_{n=0}^{\infty} A_{2n+1,0}
\frac{\sqrt{(2n)!}}{2^n n!} \sqrt{2n+1}
\end{equation}
which can be further averaged over time. 
Actual computation requires cutoffs in $n$ and $m$. Results presented
below are checked to be independent on the chosen cutoff values.
For the following results the cutoff values were $n_{max}=60$
and $m_{max}=30$.

The second approach (method II) is based on the method of characteristics. 
Taken an
equilibrium distribution at zero time, the solution of the linear
kinetic equation with
the constant relaxation time can be expressed through the equilibrium
distribution function and a statistically weighted contribution of the
trajectories in the phase space \cite{Kamke}:
\begin{eqnarray} \nonumber
f(x,p,t)  &=&  F \Bigl[\eta^{(0)}_x(t,t_0,x,p),
                           \eta^{(0)}_p(t,t_0,x,p)
                     \Bigr ] e^{-\nu (t-t_0)}  +   \\          
      &+&   \nu   \int_{t_0}^t e^{-\nu ( t - t' ) }
                       F     \Bigl[\eta_x(t',x,p),
\label{char-s}              \eta_p(t',x,p)
                             \Bigr] {\rm d } t' \;\; .
\end{eqnarray}
We remind that the functions $ \, \eta^{(0)}_{x,p} \, $ stand for the
inverted equation of a trajectory, i.e., 
they denote the dependence of the point on a trajectory
$ \,
\{x, p, t \} \, $ on its initial conditions; characteristics $ \,
\eta_{x,p} \, $
correspond to the trajectory which passes through the point $ \, \{x, p,
t' \} \, $
in the phase space. For brevity we put also $ \, t_0 = 0 $. Since
the underlying dynamical system
is non-integrable we have to find the characteristics numerically. We
choose a large
random ensemble of initial conditions with $ \, |p_0| < |p_{max}| \, $
and use
them directly to generate an approximation of $ \, f(x,p,t) \, $ in
accordance with
Eq.(\ref{char-s}). The accuracy of the approximation is controlled by
the
number of trajectories (initial conditions) used. 
We used up to $ \, 10^5 \, $ trajectories to gain good
averaging statistics. The net error of calculations was estimated as $
\, \sim 10-20\% \, $ for large relaxation parameter $\nu$ or 
large dc current $ \, j_{dc} $ after averaging of the current $j(t)$ over
several periods $T$ of the field $E(t)$. 
The maximal error has been detected 
in those cases when the average dc current tends to zero, for example, when
we put $\nu =0.001$ and $E_2=0$ (see inset of Fig. 2). In this case 
the fluctuations of the mean value of the current averaged over several
periods $T$  were of the order of residual value 
of the dc current ($\Delta j_{dc} \sim |j_{dc}|=0.002$ at $\nu=0.001$ and
$E_2=0$). In order to achieve the accuracy of the 
order of $\Delta j_{dc} /j_{dc} \sim 1/3 \div 1/2$ 
we had to perform averaging over hundreds of periods $T$. 
Note that method II has some analogy to the direct sampling
in classical Monte-Carlo simulations.

Method I is an optimal tool for not too small
relaxation $ \, \nu \ge 0.01 $ while method II shows up with larger
errors in calculations but covers, nevertheless, the limit of $ \nu \to
10^{-3} $.  Both methods give nearly identical results for $ \nu
\ge 10^{-2} \div 10^{-1} $. 

\begin{figure}[htb]
\vspace{4.0pt}
\centerline{\psfig{file=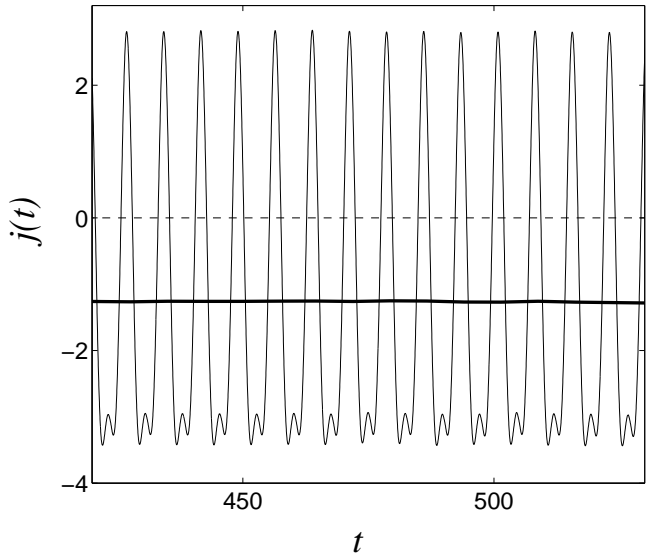,width=3.4in,angle=0}}
\vspace{3.5pt}
\caption{
Time dependence of $j(t)$ on the attractor of the kinetic equation
(thin solid line). The thick solid horizontal line shows the value of the dc part
of $j(t)$.
Parameters: $\nu=0.01$, 
$\alpha=1.5$, $v_2=0$, $U_0=6$, $E_1=-2.6$, $E_2=-2.04$, $\omega=0.85$.
The dashed line denotes the reference of $j=0$. 
}
\label{fig1-new}
\end{figure}

Fig. \ref{fig1-new} displays the dependence of the full time-dependent
current on time on the attractor of the kinetic equation computed
with the help of method II for $\nu=0.01$. We observe the expected periodic
variation in time, with a nonzero time averaged dc current indicated
by the thick horizontal line. Note that in this case the dc current
is of the order of the time-dependent current amplitude (only three
times less). This implies that the underdamped case may serve
as an ideal testing ground in various applications, since the
rectification effect is of the order of 30 percent.

\begin{figure}[htb]
\vspace{0.50pt}
\centerline{
\psfig{file=
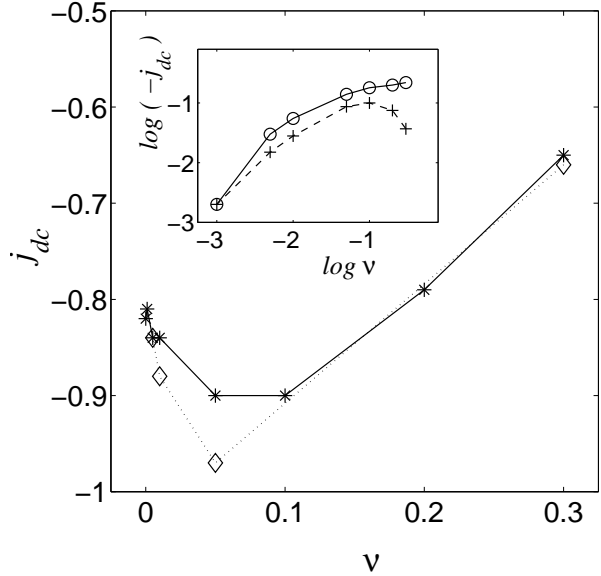,width=3.1in,angle=0}}
\vspace{3.10pt}
\caption{
Dependence of the averaged dc current (\ref{jdc}), on the dissipation
parameter  $\nu$, computed for the case of broken time symmetry
using numerical methods I ($\diamond$) and II
($\ast$) with parameters as in Fig.\ref{fig1-new}
except
$\alpha=1$.
The inset shows a
{\it log-log} plot of this dependence with parameters as in
Fig.\ref{fig1-new} except $\alpha =0$ ($\circ$) and
$E_2=0$, $v_2=0.8$, $\theta=2$ ($+$).
}
\label{fig1}
\end{figure}

In Fig. \ref{fig1} we show the dependence of the averaged dc current
on the dissipation parameter $\nu$. Results from both above 
described methods are presented. They show
data which are similar qualitatively and quantitatively
(see the caption for details).
Consider the case of $E_2\neq 0$, $\alpha \neq 0$, $v_2=0$. Directed
current appears
due to a violation of the ${\bf {\hat S}_a}$ symmetry. When 
the dissipation
$\nu$ reaches zero, this symmetry is still broken, however, the
dynamical
equations of motion may restore a second symmetry, ${\bf {\hat
S}_b}$.
But for the above set of parameters, ${\bf {\hat S}_b}$ is also
violated,
and a non-zero dc current should persist down to zero dissipation.
This is observed in Fig.
\ref{fig1}.
Suppose now we choose $\alpha=0$, then  ${\bf {\hat S}_a}$ is still
violated,
and $j_{dc} \neq 0$ for non-zero dissipation (see inset of Fig.
\ref{fig1}).
However, in the limit of zero dissipation 
${\bf {\hat S}_b}$ is restored:
$E(t)=E(-t)$ and our numerical simulations show that current tends to
zero as predicted by symmetry considerations.
The same happens in the case of a ``ratchet potential'', $v_2 \neq 0$
and without harmonic mixing ($E_2=0$). For $\nu \neq 0$
both ${\bf {\hat S}_a}$ and ${\bf {\hat S}_b}$ are violated, 
resulting in a nonzero dc current.
Again the limit of zero dissipation leads to a restoring of
${\bf {\hat S}_b}$. This leads to a disappearance
of the average dc current at $\nu=0$ (see inset of Fig. \ref{fig1}).
Thus the numerical simulations confirm our expectation 
that the attractor properties of the kinetic equation reflect the 
symmetries ${\bf {\hat S}_a}, {\bf {\hat S}_b}$ of the
underlying dynamical equations for all $\nu$ including $\nu \rightarrow 0$.

\begin{figure}[htb]
\vspace{4.0pt}
\centerline{
\psfig{file=
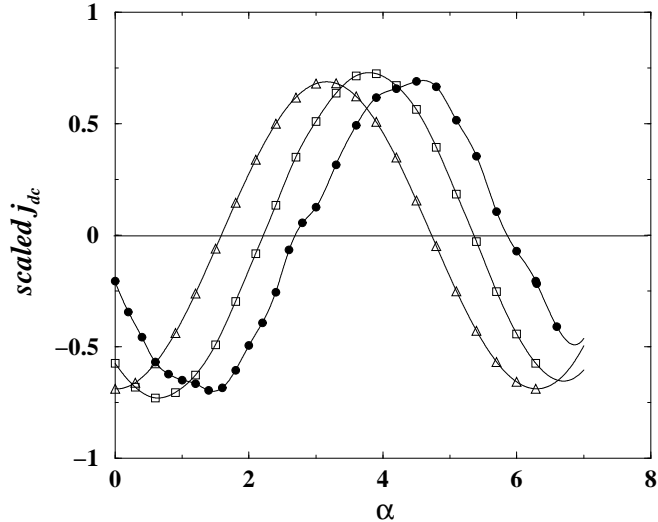,width=3.4in,angle=0}}
\vspace{3.5pt}
\caption{
Dependence of the averaged dc current on the phase delay parameter
$\alpha$
for different values of $\nu$: $\nu=0.3$ (filled circles),
$\nu=1$ (squares) and $\nu=4$ (triangles). Note that the
current values are scaled by a factor of 4.86 ($\nu=1$) and
181.6 ($\nu=4$). The lines connecting the data are splines.
Other parameters are as in Fig.\ref{fig1-new}.
}
\label{fig2}
\end{figure}

In Fig. \ref{fig2} we show the dependence of the averaged dc current on
the
``phase difference'' $\alpha$ for three different values of the
dissipation
parameter $\nu=4,1,0.3$, computed with the help of method I. 
Note that the currents for $\nu=4$ and $\nu=1$ are scaled with
factors 181.6 and 4.86 respectively. consequently we observe
an enhancement of the maximum dc current value by more than two
orders of magnitude when going from the overdamped case ($\nu=4$)
to the underdamped one ($\nu=0.3$). Besides this observation 
Fig. \ref{fig2} displays again the consequences of symmetry breaking
or restoral. With increasing $\nu$ the dc current becomes zero
for $\alpha = \pm \pi/2$ which is a consequence of restoring of
${\bf {\hat S}_c}$ symmetry. With decreasing $\nu$ we observe
a vanishing of the current at $\alpha =0,\pi$ which is a consequence
of restoring of the ${\bf {\hat S}_b}$ symmetry. 

{\it Conclusions.} In this paper, we extend the symmetry approach
for explanation of a dc current in driven nonlinear systems
to
the framework of the classical kinetic Boltzmann equation. We have
analysed separately the cases of large and small dissipation and have
demonstrated that the rectification of current due to nonlinear mixing
of harmonics takes place in agreement with previous
symmetry consideration \cite{fyz00prl}. In particular, it follows that
in the limit
of small $ \, \nu \, $ the attractor of the Boltzmann
equation (\ref{KE-1}-\ref{KE-2})
reflects the dynamical symmetries of the Newtonian equations 
of motion\cite{comm1}.
Finally we observe 
that a proper choice of the system parameters
in the underdamped case ensures the value of dc current
to be of the order of the ac current amplitude. The underdamped
case leads to an enhancement of the dc current by several orders 
of magnitude as compared to the overdamped one.

The authors wish to thank T. Dittrich, M. Fistul, I. Goychuk and H. Schanz 
for helpful and stimulating discussions. We also would like to thank
P. Reimann for sending us preprints prior to publication.



\begin{thebibliography}{0}
\bibitem{comm1} Note that the consideration of a tight-binding model
\cite{ackkc98} gives zero current in the $\nu \rightarrow 0$ limit and
$\cos {\alpha}$ dependence [cf. (16)] for all $\nu$ due to
the presence of an additional symmetry of the $\nu=0$ case. Details
will be published elsewhere.
\end{thebibliography}
\end{document}